\documentclass[12pt]{article}

\usepackage[utf8]{inputenc}
\usepackage[utf8]{inputenc}

\usepackage{amssymb}
\usepackage{amsfonts}
\usepackage{graphicx}
\usepackage{circuitikz}
\usetikzlibrary{quantikz}

\usepackage[verbose]{placeins} 

\usepackage{braket}
\usepackage{tikz}
\usepackage[T1]{fontenc}
\usepackage{imakeidx}

\usepackage{xcolor,varwidth}

\usepackage{pgfplots}
 \usetikzlibrary{backgrounds,calc}
 \usepackage{amsmath,tikz}
 \usetikzlibrary{quotes,angles}
\pgfplotsset{compat = newest}
\usetikzlibrary{quotes,angles}

\usetikzlibrary{svg.path}
\usepackage{pgfornament}

\usepackage{xcolor}

\usepackage{subfig}
\usepackage{graphicx}
\def\mset #1[#2]=#3{%
	\expandafter\xdef\csname #1#2\endcsname{#3}
}
\def\mget #1[#2]{%
	\csname #1#2\endcsname
}
\def\minc #1[#2]+=#3{%
	\pgfmathparse{\mget #1[#2]+#3}%
	\mset #1[#2]=\pgfmathresult
}

\date{}

\usepackage{hyperref}
\usepackage [all]{hypcap}
\usepackage{color}

\usepackage{listings}
\definecolor{codegreen}{rgb}{0,0.6,0}
\definecolor{codegray}{rgb}{0.5,0.5,0.5}
\definecolor{codepurple}{rgb}{0.58,0,0.82}
\definecolor{backcolour}{rgb}{0.95,0.95,0.92}

\lstdefinestyle{mystyle}{
    backgroundcolor=\color{backcolour},   
    commentstyle=\color{codegreen},
    keywordstyle=\color{magenta},
    numberstyle=\tiny\color{codegray},
    stringstyle=\color{codepurple},
    basicstyle=\ttfamily\footnotesize,
    breakatwhitespace=false,         
    breaklines=true,                 
    captionpos=b,                    
    keepspaces=true,                 
    numbers=left,                    
    numbersep=5pt,                  
    showspaces=false,                
    showstringspaces=false,
    showtabs=false,                  
    tabsize=2
}

\begin{document}

\title{ Area and volume as emergent phenomena from entangled qubits}
\author{ Juan M. Romero \thanks{jromero@cua.uam.mx } and  Emiliano Montoya-Gonz\'alez \thanks{emiliano.montoya.g@cua.uam.mx }    \\
Departamento de Matemáticas Aplicadas y Sistemas,\\
Universidad Aut\'onoma Metropolitana-Cuajimalpa,\\
M\'exico, D.F 05300, M\'exico. }

\date{\today}

\maketitle
\begin{abstract}

Recently, a connection has been shown between certain geometric quantities and quantum information theory. In this paper, we demonstrate that geometric quantities such as area and volume can emerge directly from entangled multi-qubit states. In particular, the area of a two-dimensional parallelogram is derived from a 4-qubit entangled state, the vector area of a three-dimensional parallelogram from three 6-qubit entangled states, and the volume of a three-dimensional parallelepiped from a 9-qubit entangled state. Corresponding quantum circuits are constructed and implemented using Qiskit to generate the required entangled states. Given that parallelograms and parallelepipeds serve as elementary building blocks for more complex geometric structures, these results may offer a pathway toward exploring emergent geometry in quantum information frameworks.\\


\end{abstract}

\section{\label{sec:introduction}Introduction}

One of the major problems in physics is the quantization of general relativity. Since general relativity is a geometric theory, developing 
a quantum version of its geometric structures is important for constructing a theory of quantum gravity.
 In this context a remarkable result is the connection between certain geometric quantities and concepts from quantum 
information theory.  For instance, in the holographic correspondence, it has been shown that the entanglement entropy of certain quantum field theories is related to the area of the event horizon of black holes  \cite{Ryu,Van,Van1}. Moreover,  recent theoretical  proposals a correspondence between the geometry of  Einstein-Rosen bridges, commonly referred to as wormholes, and the maximally entangled states of two black holes, the so-called  $EPR=ER$ conjecture  \cite{Malda,Malda1,Malda2}. Additionally, quantum computing frameworks enable the extraction of 
geometries properties- such as spacetime metrics- from quantum circuit complexity, some of  which are associated with black-holes geometries \cite{Brown,Brown1,Haferkamp,Chapman}. Based on these findings, several authors have suggested that space time itself may emerge from entangled states.\\

In this paper, it is shown that the area of a parallelogram and the volume of a parallelepiped can be derived directly from entangled qubits. To obtain these results, the area and volume are expressed as multilinear polynomials. The area of a $2D$ parallelogram is derived from an entangled $4$-qubit state. Furthermore, the vector area of a $3D$ parallelogram is obtained from three entangled $6$-qubit states, and the volume of a $3D$ parallelepiped is deduced from an entangled $9$-qubit state. The corresponding quantum circuits for these entangled states are provided in Qiskit code.\\

It is worth noting that parallelograms and parallelepipeds serve as fundamental building blocks for more complex geometric structures. Therefore, the results presented in this paper may facilitate the study of more sophisticated geometric systems using quantum methods.\\

This paper is organized as follows. In Sec. \ref{Poly}, a  brief review  of the  relation between multilinear  polynomials and states of qubit is provide. In Sec. \ref{2DArea}, the  area of a $2D$ parallelogram  is  obtained from an entangled  $4$-qubit state. In Sec. \ref{3DArea}, the vector area of a $3D$ parallelogram  is deduced from  three  entangled state.  In Sec. \ref{3DVolume}, the  volume of a $3D$  parallelepiped  is derived from an entangled state.  Finally, in Sec. \ref{Con}, a summary is given.

\section{Multilinear Polynomials}
\label{Poly}
 
 In this section, multilinear polynomials and their relation to qubit states  are reviewed.\\

Firts, let us note that if 
\begin{eqnarray}
c_{1}, c_{2}, c_{3},c_{4}\label{c1}
\end{eqnarray}
are complex constants, the following  bilinear polynomial  can be constructed
\begin{eqnarray}
 P(x,y)=c_{1}+c_{2}x+c_{3}y+c_{4}xy.  \label{bl1}
\end{eqnarray}
In addition,  in a two-qubit system, by  using the constants  \eqref{c1}  the state
\begin{eqnarray}
\ket{\psi}=c_{0}\ket{00}+c_{1}\ket{01}
+c_{2}\ket{10}+c_{12}\ket{11}.\label{c1A1}
\end{eqnarray}
can be constructed.
Then, for each bilinear polynomial \eqref{bl1} a two-qubit state \eqref{c1A1} can be constructed. 
In particular, a non-factorable  bilinear polynomial  can be  related to entangled two qubit. In fact,  teleportation can be modelled with multilinear polynomials \cite{Yo}.\\

Notice that if the equation  
\begin{eqnarray}
|c_{1}|^{2}+| c_{2}|^{2}+|c_{3}|^{2}+|c_{4}|^{2}=1\nonumber
\end{eqnarray}
is satisfied, the state \eqref{c1A1} is normalized. \\

 In general, a multilinear polynomials of  $n$ variables can be written as sum of  $2^{n}$ terms, like
\begin{eqnarray}
  & &P(x_{1},\cdots,x_{n}) =c_{0}+\sum_{i=1}^{n}c_{i}x_{i}+\sum_{1\leq i_{1}<i_{2}\leq n }c_{i_{1}i_{2}}x_{i_{1}}x_{i_{2}}+ \cdots+\nonumber\\
 & & +\sum_{1\leq i_{1}<i_{2}<\cdots<i_{k}\leq n }c_{i_{1}i_{2} \cdots i_{k}} x_{i_{1}}x_{i_{2}} \cdots x_{i_{k}}+\cdots+  c_{123 \cdots n } x_{1}x_{2}x_{3} \cdots x_{n}. 
 \label{bln}
\end{eqnarray}

Moreover, a basis of  system of $n$ qubits must have  $2^{n}$ states. Thus, in this system every state $\ket{\psi}$ is a linear combination of $2^{n}$ states.
In particular, by using the constants of the multilinear polynomial \eqref{bln}, the following state can be constructed 
\begin{eqnarray}
 \ket{\psi}&=&c_{0}\ket{00\cdots 0}+c_{1}\ket{0\cdots 01}+c_{2}\ket{0\cdots 10}+\cdots+c_{n}\ket{10\cdots0}+ \nonumber\\
               & &+c_{12}\ket{0\cdots0 11}+c_{13}\ket{0\cdots01 01}+\cdots c_{n-1n}\ket{110\cdots0}+\cdots+\nonumber\\
 & &+c_{123\cdots n}\ket{11\cdots 1}.  \label{c1An}
\end{eqnarray}
Therefore, each multilinear polynomial \eqref{bln}  is related to  a $n$-qubit state \eqref{c1An}.

\section{Scalar area and  entangled qubits}
\label{2DArea}
The area of a $2D$ parallelogram   can be expressed as a multilinear polynomials. 
For example,  if a  parallelogram is  built  with the  $2D$ dimensional vectors 
\begin{eqnarray}
v_{1}&=&(x_{1},x_{2}), \nonumber \\
v_{2}&=&(x_{3},x_{4})  \nonumber
\end{eqnarray}
the area is given by 
\begin{eqnarray}
A(v_{1},v_{2})=\det \begin{pmatrix} x_{1}&x_{2}\\ x_{3}& x_{4}\end{pmatrix}=x_{1}x_{4}-x_{2}x_{3}. \nonumber
\end{eqnarray}
Namely, the area of parallelogram  is given by   multilinear polynomials of $4$ variables 
\begin{eqnarray}
A(v_{1},v_{2})=x_{1}x_{4}-x_{2}x_{3}. \label{Area2DE}
\end{eqnarray}

Now, a basis of a system of $4$ qubits must have $2^{4}=16$ states. Consequently, a general state of this system can be written as follows:
\begin{eqnarray}
\ket{\psi}&=&a_{0}\ket{0000}+a_{1}\ket{0001}+a_{2} \ket{0010}+a_{3}\ket{0100}+a_{4}\ket{1000}+a_{12}\ket{0011}+\nonumber\\
& &+a_{13}\ket{0101}+a_{14}\ket{1001}+a_{23}\ket{0110}+a_{24}\ket{1010}+a_{34}\ket{1100}+\nonumber\\
&& +a_{123}\ket{0111}+a_{124}\ket{1011}+a_{134}\ket{1101}+a_{234}\ket{1110}+a_{1234}\ket{1111}.  \nonumber
\end{eqnarray}
A particular case is given by the entangled state
\begin{eqnarray}
\ket{A}=\ket{1001}-\ket{0110}. \label{area2D}
\end{eqnarray}

Notice that by using the state \eqref{area2D}, the  expression
\begin{eqnarray}
\braket{A|\psi}=a_{14}-a_{23}  \nonumber
\end{eqnarray}
is found. Then, if the  equations
\begin{eqnarray}
a_{14}&=&x_{1}x_{4},  \nonumber\\ 
a_{23}&=&x_{2}x_{3} \nonumber
\end{eqnarray}
are  satisfied, the area of parallelogram  \eqref{Area2DE} is obtained. Namely
\begin{eqnarray}
{\rm Area}=\braket{A|\psi}=x_{1}x_{4}-x_{2}x_{3}.  \nonumber
\end{eqnarray}
 Then, the area of a $2D$ parallelogram is obtained from an entangled $4$-qubit state.\\

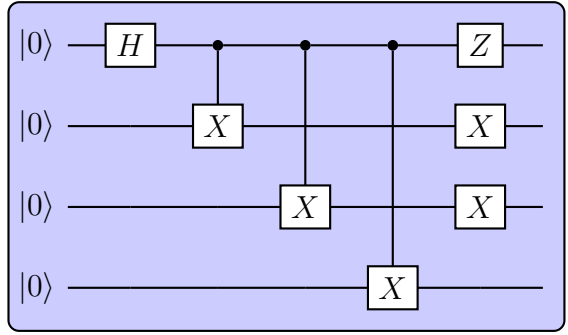
\begin{figure}[hbt!]
\centering
\begin{quantikz}
\gategroup[wires=4,steps=8,style={rounded corners,fill=blue!20}, background]{}
&\lstick{$|0\rangle$} & \gate{H} & \ctrl{1} & \ctrl{2} & \ctrl{3}     & \gate{Z} & \qw \\
&\lstick{$|0\rangle$} & \qw  & \gate{X}  & \qw      & \qw                    & \gate{X} & \qw   \\
&\lstick{$|0\rangle$} & \qw      & \qw      & \gate{X}  & \qw                 & \gate{X} & \qw      \\
&\lstick{$|0\rangle$} & \qw      & \qw      & \qw      & \gate{X}  &   \qw                   & \qw
\end{quantikz}
\caption{Quantum circuit for the entangled states \eqref{area2D}. }
    \label{fig:CArea2D}
\end{figure}

The Qiskit code for  the entangled state \eqref{area2D} is given in Listing  \ref{lst:A2D} and their  quantum circuit is in the Figure \ref{fig:CArea2D}.\\

\begin{lstlisting}[language=Python, caption=Qiskit code for  the entangled  state \eqref{area2D}.,  label={lst:A2D} ]
from qiskit import QuantumCircuit
from qiskit.quantum_info import Statevector
from qiskit.visualization import plot_state_qsphere
circuit = QuantumCircuit(4)
circuit.h(0)
circuit.cx(0, 1)
circuit.cx(0, 2)
circuit.cx(0, 3)
circuit.barrier()
circuit.x(1)
circuit.x(2)
circuit.barrier()
circuit.z(0)
\end{lstlisting}

\section{Vector area and entangled qubits}
\label{3DArea}

Now, the area vector  for two $3D$  vectors 
\begin{eqnarray}
v_{1}&=&(x_{1},x_{2},x_{3}),  \nonumber \\
v_{2}&=&(x_{4}, x_{5},x_{6})   \nonumber
\end{eqnarray}
is given by 
\begin{eqnarray}
A(v_{1},v_{2})&=&\det \begin{vmatrix} \hat i& \hat j&\hat k\\x_{1}&x_{2}&x_{3}\\x_{4}&x_{5}&x_{6}  \end{vmatrix}=\nonumber\\
&=&\left(  x_{2}x_{6}-x_{3}x_{5}\right)\hat i+\left( x_{3}x_{4}-x_{1}x_{6}\right)\hat j+\left( x_{1}x_{5}-x_{2}x_{4}\right)\hat k. \label{varea}
\end{eqnarray}
In this case, we have three multilinear polynomials with $6$ variables 
\begin{eqnarray}
A_{1}&=&  x_{2}x_{6}-x_{3}x_{5},\nonumber\\ 
A_{2}&=& x_{3}x_{4}-x_{1}x_{6}, \nonumber\\
A_{3}&=& x_{1}x_{5}-x_{2}x_{4}\nonumber.
\end{eqnarray}

Moreover, a basis of a system of $6$ qubits must have $2^{6}=64$ states. Then, a general state of this system can be written  as follows:
\begin{eqnarray}
\ket{\psi}&=&a_{0}\ket{000000}+a_{1}\ket{000001}+a_{2} \ket{000010}+a_{3}\ket{000100}+a_{4}\ket{001000}+a_{5}\ket{010000}\nonumber\\
& &+a_{6}\ket{100000}+a_{12}\ket{000011}+a_{13}\ket{000101}+\cdots +a_{123456}\ket{111 111}.  \nonumber
\end{eqnarray}
Notice that in this system the following entangled states 
\begin{eqnarray}
\ket{A_{1}}&=&\ket{100010}-\ket{010100}, \label{area3D1}\\
\ket{A_{2}}&=&\ket{001100}-\ket{100001}, \label{area3D2} \\
\ket{A_{3}}&=&\ket{010001}-\ket{001010} \label{area3D3}  
\end{eqnarray}
can be constructed.\\

Then by using the states \eqref{area3D1}-\eqref{area3D3}, the equations
\begin{eqnarray}
\braket{A_{1}|\psi}=a_{26}-a_{35},  \nonumber\\ 
\braket{A_{2}|\psi}=a_{34}-a_{16},  \nonumber\\ 
\braket{A_{3}|\psi}=a_{15}-a_{24}  \nonumber
\end{eqnarray}
are found.\\

Therefore, if 
\begin{eqnarray}
a_{26}&=&x_{2}x_{6},\nonumber \\ 
a_{35}&=&x_{3}x_{5}, \nonumber\\
a_{34}&=&x_{3}x_{4},  \nonumber\\ 
a_{16}&=&x_{1}x_{6},  \nonumber\\
a_{15}&=&x_{1}x_{5},   \nonumber\\ 
a_{24}&=&x_{2}x_{4}  \nonumber
\end{eqnarray}
the  expressions 
\begin{eqnarray}
{\rm Area}_{1}&=&\braket{A_{1}|\psi}=x_{2}x_{6}-x_{3}x_{5},  \nonumber\\
{\rm Area}_{2}&=&\braket{A_{2}|\psi}=x_{3}x_{4}-x_{1}x_{6},  \nonumber\\
{\rm Area}_{3}&=&\braket{A_{3}|\psi}=x_{1}x_{5}-x_{2}x_{4}  \nonumber
\end{eqnarray}
are gotten, which are the components of the area vector \eqref{varea}. \\

 Thus, the vector area of a $3D$ parallelogram is derived from three entangled 6-qubit states.\\

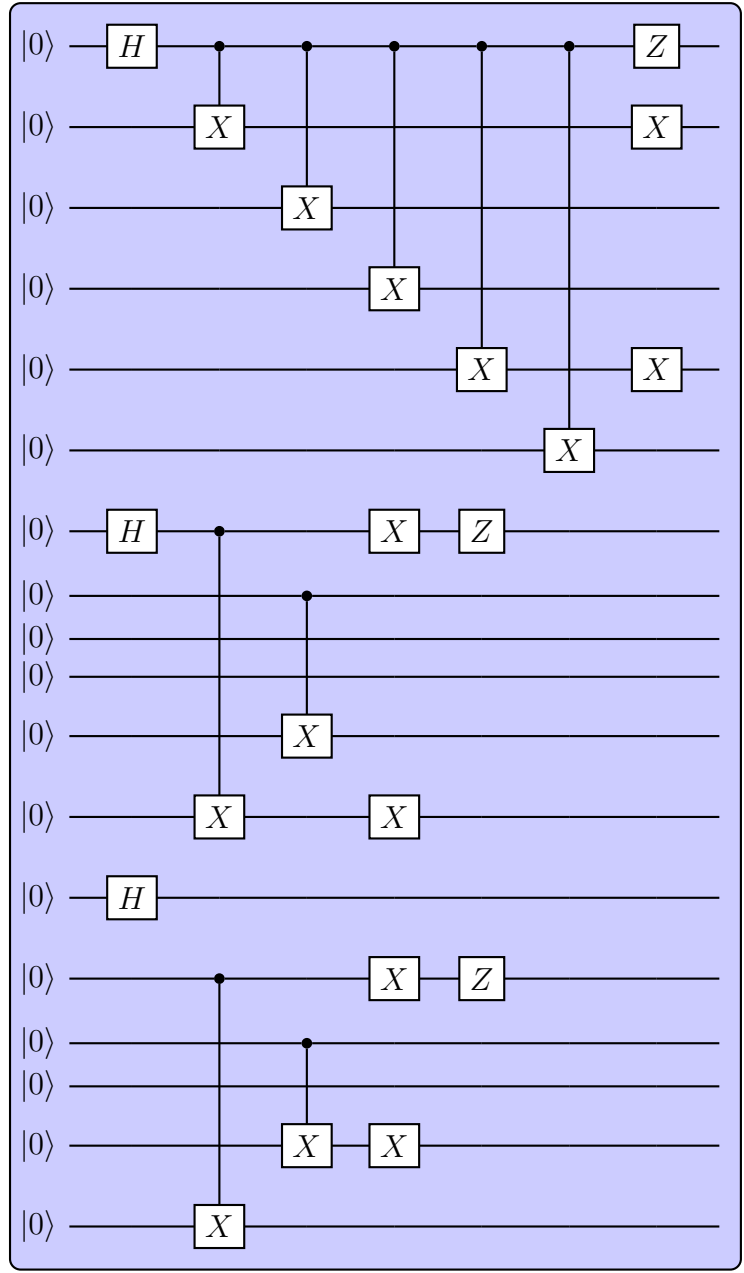
\begin{figure}[hbt!]
\centering
\begin{quantikz}
\gategroup[wires=18,steps=10,style={rounded corners,fill=blue!20}, background]{}
&\lstick{$|{0}\rangle$} & \gate{H} &\ctrl{1} &\ctrl{2} &\ctrl{3} &\ctrl{4} &\ctrl{5} & \gate{Z} & \qw
\\
&\lstick{$|{0}\rangle$}  & \qw & \gate{X} & \qw & \qw & \qw & \qw & \gate{X} & \qw \\
&\lstick{$|{0}\rangle$}  & \qw & \qw  & \gate{X} & \qw & \qw & \qw & \qw & \qw \\
&\lstick{$|{0}\rangle$}  & \qw & \qw  & \qw & \gate{X} & \qw & \qw & \qw & \qw \\
&\lstick{$|{0}\rangle$}  & \qw & \qw  & \qw & \qw & \gate{X} & \qw & \gate{X} & \qw \\
&\lstick{$|{0}\rangle$}  & \qw & \qw & \qw & \qw & \qw & \gate{X} & \qw & \qw \\ 
&\lstick{$|{0}\rangle$}  & \gate{H} & \ctrl{5} & \qw & \gate{X} & \gate{Z} & \qw & \qw & \qw \\
&\lstick{$|{0}\rangle$}  & \qw & \qw & \ctrl{3} & \qw & \qw & \qw & \qw & \qw \\
&\lstick{$|{0}\rangle$}  & \qw & \qw & \qw & \qw & \qw & \qw & \qw & \qw \\
&\lstick{$|{0}\rangle$}  & \qw & \qw & \qw & \qw & \qw & \qw & \qw & \qw \\
&\lstick{$|{0}\rangle$}  & \qw & \qw & \gate{X} & \qw & \qw & \qw & \qw & \qw \\
&\lstick{$|{0}\rangle$}  & \qw & \gate{X} & \qw & \gate{X} & \qw & \qw & \qw & \qw \\ 
&\lstick{$|{0}\rangle$}  & \gate{H} & \qw & \qw & \qw & \qw & \qw & \qw & \qw \\
&\lstick{$|{0}\rangle$}  & \qw & \ctrl{4} & \qw & \gate{X} & \gate{Z} & \qw & \qw & \qw \\
&\lstick{$|{0}\rangle$}  & \qw & \qw & \ctrl{2} & \qw & \qw & \qw & \qw & \qw \\
&\lstick{$|{0}\rangle$}  & \qw & \qw & \qw & \qw & \qw & \qw & \qw & \qw \\
&\lstick{$|{0}\rangle$}  & \qw & \qw & \gate{X} & \gate{X} & \qw & \qw & \qw & \qw \\
&\lstick{$|{0}\rangle$}  & \qw & \gate{X} & \qw & \qw & \qw & \qw & \qw & \qw 
\end{quantikz}
\caption{Quantum circuit for the entangled states  \eqref{area3D1}-\eqref{area3D3}. }
    \label{fig:CArea3D}
\end{figure}

The Qiskit code for  the entangled states \eqref{vec3D1}-\eqref{vec3D3}  is given in Listing \ref{lst:A3D} and their  quantum circuit is in the Figure \ref{fig:CArea3D}.\\ \\

\begin{lstlisting}[language=Python, caption=Qiskit code for  the entangled  states \eqref{area3D1}-\eqref{area3D3}.,  label={lst:A3D} ]
from qiskit import QuantumRegister, ClassicalRegister, QuantumCircuit
from qiskit_aer import AerSimulator


qreg_q = QuantumRegister(18, 'q')
creg_c = ClassicalRegister(18, 'c')
circuit = QuantumCircuit(qreg_q, creg_c)


circuit.h(qreg_q[0])  
circuit.barrier()


circuit.cx(qreg_q[0], qreg_q[1])
circuit.cx(qreg_q[0], qreg_q[2])
circuit.cx(qreg_q[0], qreg_q[3])
circuit.cx(qreg_q[0], qreg_q[4])
circuit.cx(qreg_q[0], qreg_q[5])
circuit.barrier()


circuit.x(qreg_q[1])
circuit.x(qreg_q[4])
circuit.z(qreg_q[0])
circuit.barrier()


circuit.h(qreg_q[6])  
circuit.barrier()


circuit.cx(qreg_q[6], qreg_q[11])
circuit.cx(qreg_q[7], qreg_q[10])
circuit.barrier()


circuit.x(qreg_q[6])
circuit.x(qreg_q[11])
circuit.z(qreg_q[6])
circuit.barrier()


circuit.h(qreg_q[12])  
circuit.barrier()


circuit.cx(qreg_q[13], qreg_q[17])
circuit.cx(qreg_q[14], qreg_q[16])
circuit.barrier()


circuit.x(qreg_q[13])
circuit.x(qreg_q[16])
circuit.z(qreg_q[13])
circuit.barrier()

print(circuit)
\end{lstlisting}

\section{Volume and entangled qubits}
\label{3DVolume}

Additionally, the volume of a three-dimensional  parallelepiped can be expressed as a multilinear polynomials. 
In fact,  for a  parallelepiped built  with the following  $3D$ vectors 
\begin{eqnarray}
v_{1}&=&(x_{1},x_{2},x_{3}), \label{vec3D1}\\
v_{2}&=&(x_{4}, x_{5},x_{6}), \label{vec3D2}\\
v_{3}&=&(x_{7},x_{8},x_{9})\label{vec3D3}
\end{eqnarray}
the volume  is given by 
\begin{eqnarray}
V(v_{1},v_{2},v_{3})&=&\det \begin{pmatrix} x_{1}&x_{2}&x_{3}\\x_{4}&x_{5}&x_{6}\\ x_{7}&x_{8}&x_{9}  \end{pmatrix}=\nonumber\\
&=&x_{1}x_{5}x_{9}-x_{1}x_{6}x_{8}+  x_{2}x_{6}x_{7}-x_{2}x_{4}x_{9}+x_{3}x_{4}x_{8}-x_{3}x_{5}x_{7}. \nonumber
\end{eqnarray}  
That is, the volume  of  parallelepiped  is given by a  multilinear polynomials of $9$ variables 
\begin{eqnarray}
V(v_{1},v_{2},v_{3})=x_{1}x_{5}x_{9}-x_{1}x_{6}x_{8}+  x_{2}x_{6}x_{7}-x_{2}x_{4}x_{9}+x_{3}x_{4}x_{8}-x_{3}x_{5}x_{7}. \label{volumep}
\end{eqnarray}

Furthermore, a basis of a system of $9$ qubits must have $2^{9}=512$  states. Thus, a general state of this system  can be written as follows:
\begin{eqnarray}
\ket{\psi}&=&a_{0}\ket{ 0 00 00 00 00}+a_{1}\ket{0 00 00 00 01}+a_{2} \ket{0 00 00 0010}+a_{3}\ket{0 00 00 0100} +\nonumber\\
& &+a_{4}\ket{0 00 00 1000}+a_{5}\ket{0  00 01 00 00}+a_{6}\ket{0 00 10 00 00}+a_{7}\ket{0 01 00 00 00}+\nonumber\\
& &+a_{8}\ket{0  10 00 00 00}+a_{9}\ket{100000000}+a_{12}\ket{0 00 00 00  11}+\cdots +\nonumber\\
& &+a_{123456789}\ket{1 11 11 11  11}.  \nonumber
\end{eqnarray}

In particular,  the  entangled state
\begin{eqnarray}
\ket{V}&=&\ket{100010001}-\ket{010100001}+\ket{001100010} -\ket{100001010}+\nonumber\\
& &+\ket{010001100}-\ket{001010100} \label{volume}
\end{eqnarray}
can be constructed. \\

Thus,  by using the state \eqref{volume}, the equation
\begin{eqnarray}
\braket{V|\psi}=a_{159}-a_{168}+a_{267}-a_{249}+a_{348}-a_{357} \nonumber
\end{eqnarray}
is obtained.\\ 

Then, when expressions 
\begin{eqnarray}
a_{159}&=&x_{1}x_{5}x_{9},\nonumber\\
  a_{168}&=&x_{1}x_{6}x_{8,}\nonumber\\ 
  a_{267}&=&x_{2}x_{6}x_{7},\nonumber\\
a_{249}&=&x_{2}x_{4}x_{9},\nonumber\\ 
a_{348}&=&x_{3}x_{4}x_{8},\nonumber\\ 
a_{357}&=&x_{3}x_{5}x_{7}, \nonumber
\end{eqnarray}
are satisfied,   the volume of the  parallelepiped \eqref{volumep} 
\begin{eqnarray}
\braket{V|\psi}=V=x_{1}x_{5}x_{9}-x_{1}x_{6}x_{8}+  x_{2}x_{6}x_{7}-x_{2}x_{4}x_{9}+x_{3}x_{4}x_{8}-x_{3}x_{5}x_{7}  \nonumber
\end{eqnarray}
is gotten.\\

Hence,  the volume of a $3D$ parallelepiped is obtained from an entangled $9$-qubit state.\\

The results of this paper support the idea that the space can emerge from entangled states.\\ 

\begin{figure}[hbt!]
\centering
\begin{quantikz}
\gategroup[wires=5,steps=12,style={rounded corners,fill=blue!20}, background]{}
&\lstick{$|{0}\rangle$} & \gate{H} & \qw      & \qw      & \qw      & \qw      & \qw & \ctrl{1}      & \gate{H} & \meter{} & \qw \\
&\lstick{$|{0}\rangle$}  & \qw      & \gate{H} & \ctrl{1} & \gate{Z}     & \qw & \qw & \gate{X}  & \ctrl{1} & \meter{} & \qw  \\
&\lstick{$|{0}\rangle$} & \qw      & \gate{X} & \targ{}  & \ctrl{1} & \qw      & \qw      & \qw      & \targ{}  & \qw      & \qw  \\
&\lstick{$|{0}\rangle$}  & \qw      & \qw      & \qw      & \gate{X}  & \ctrl{1} & \qw      & \qw      & \qw      & \qw   & \qw  \\
&\lstick{$|{0}\rangle$}  & \qw      & \gate{X} & \qw      & \qw      & \targ{}  & \gate{Z} & \qw      & \qw      & \qw      & \qw  \\
\gategroup[wires=5,steps=12,style={rounded corners,fill=blue!20}, background]{}
&\lstick{...} & \qw & \qw      & \qw      & \qw  & \qw & \qw & \qw & \qw & \qw & \qw \\
&\lstick{...}  & \qw      & \qw & \qw & \qw & \qw & \qw & \qw & \qw & \qw & \qw \\
&\lstick{...} & \qw      & \qw & \qw  & \qw & \qw & \qw & \qw & \qw & \qw & \qw \\
&\lstick{...}   & \gate{If(0)\: Z} & \gate{IF(2)\: Y} & \gate{IF(3)\: X} & \qw & \qw & \qw & \qw & \qw & \qw & \qw \\
&\lstick{...}  & \qw      & \qw & \qw      & \qw   & \qw & \qw & \qw & \qw & \qw & \qw
\end{quantikz}
\caption{Quantum circuit for the entangled state  \eqref{volume}. }
    \label{fig:Vol3D}
\end{figure}
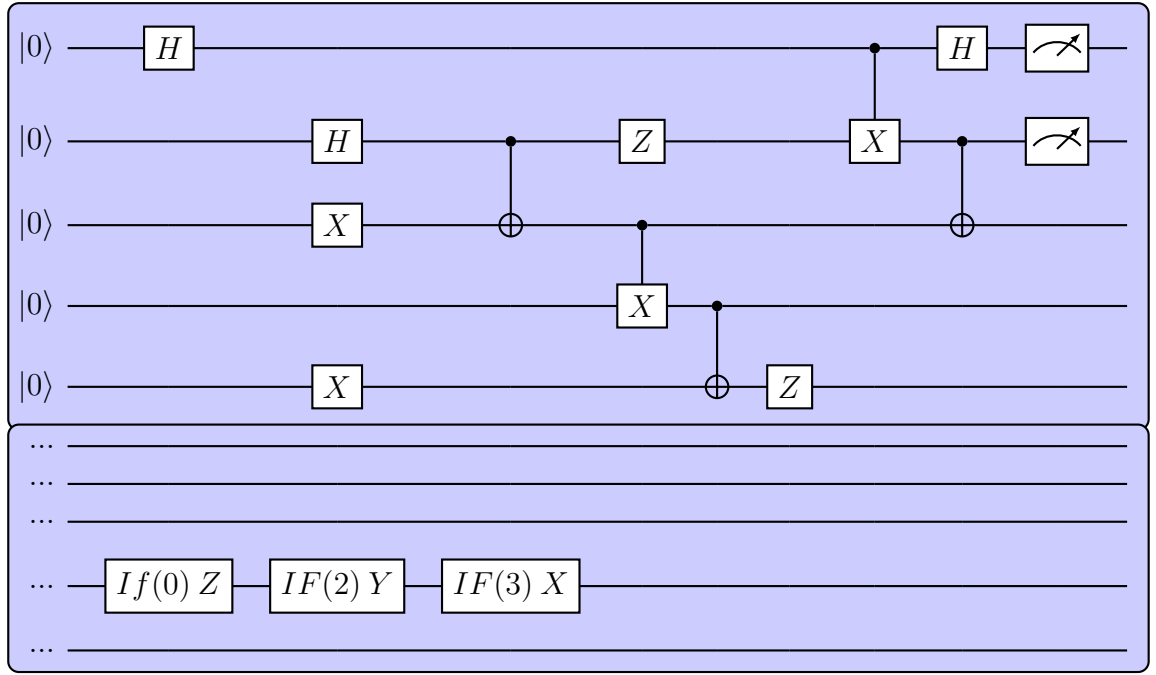

The Qiskit code for  the entangled state  \eqref{volume}  is given in Listing \ref{lst:Vol} and their  quantum circuit is in the Figure \ref{fig:Vol3D}. \\

\begin{lstlisting}[language=Python, caption=Qiskit code for  the  entangled state \eqref{volume}.,  label={lst:Vol} ]
from qiskit import QuantumRegister, ClassicalRegister, QuantumCircuit
from qiskit_aer import AerSimulator

qc = QuantumCircuit(9)


qc.h(0)
qc.h(1)
qc.h(2)


qc.x([1, 2])  
qc.ccx(0, 1, 3)  
qc.ccx(0, 2, 7)  
qc.x([1, 2])  

qc.x([0, 2])
qc.ccx(1, 0, 4)  
qc.ccx(1, 2, 6)  
qc.z(1)
qc.x([0, 2])


qc.x([0, 1])
qc.ccx(2, 0, 5)  
qc.ccx(2, 1, 6) 
qc.cx(2, 8)     
qc.x([0, 1])

qc.save_density_matrix()

print(qc)

simulator = AerSimulator()
result = simulator.run(qc).result()
state = result.data()['density_matrix']
print(state)
\end{lstlisting}

Recently, significant advancements have been made in generating entangled photon states in quantum optics, particularly through nonlinear processes \cite{3P1,3P2}. 
These results have boosted research in quantum computing and quantum communication using photons. Notably, the mathematical framework used to analyze entangled qubit states is essentially the same as that applied to entangled photon states in quantum optics. Then, the results of this paper can be used to study entangled photon states, too.

\section{Conclusions}
\label{Con}

In this paper, the area of a $2D$ parallelogram is obtained from an entangled $4$-qubit state. In addition, the vector area of a $3D$ parallelogram is derived from three entangled 6-qubit states. Moreover, the volume of a $3D$ parallelepiped is obtained from an entangled $9$-qubit state. 
Additionally, the corresponding quantum circuits for these entangled states were provided in Qiskit code.\\

The results of this paper  support the hypothesis that the spacetime  can emerge from entangled states.\\ 

It is worth mentioning that parallelograms and parallelepipeds serve as fundamental building blocks for more sophisticated geometric structures. Therefore, the results presented in this work may enable the study of more complex geometries, such as those arising in general relativity. We intend to explore this subject in future work.\\

\section*{Acknowledgements}
E. M-G. is supported by SECIHTI Master fellowship.


\begin{thebibliography}{99}

\bibitem{Ryu} 
S. Ryu and T. Takayanagi,
{\it Holographic derivation of entanglement entropy from $AdS/CFT$,} 
 Phys. Rev. Lett. {\bf 96}, 181602 (2006)
 
 \bibitem{Van} 
 M.  Van Raamsdonk, {\it  Building up spacetime with quantum entanglement,}
Int. J. Mod. Phys. D {\bf 19}, 2429 (2010).
 
 
  \bibitem{Van1} 
  R.  Bousso and S. Kaya, {\it Holographic entropy cone beyond $AdS/CFT$,}
Phys. Rev. D {\bf 111}, 086014 (2025).

 \bibitem{Malda} 
 J. Maldacena and L. Susskind, {\it Cool horizons for entangled black holes,}
Fortschr. Phys. {\bf 61}, No. 9, 781 (2013).
 
  \bibitem{Malda1}
  S. Raju, {\it Lesson from the information paradox,}  Physics Reports
{\bf 943}, 1 (2022).

  
 \bibitem{Malda2}
 A. Almheiri, T. Hartman, J. Maldacena, E. Shaghoulian, and A.  Tajdini, {\it The entropy of Hawking radiation,}
Rev. Mod. Phys. {\bf 93}, 035002 (2021).


 \bibitem{Bengtsson}   
 I.  Bengtsson and  K.  Zyczkowski, 
 {\it Geometry of quantum states: an introduction to quantum entanglement,}
 Cambridge University Press; 2nd edition,  England  (20017).

 
\bibitem{Brown}  
A. R. Brown, L. Susskind, and Y. Zhao, {\it Quantum complexity and negative curvature,}  
Phys. Rev. D {\bf 95}, 045010 (2017).

\bibitem{Brown1}  
A. R. Brown and L. Susskind, {\it Second law of quantum complexity,}
Phys. Rev. D {\bf 97}, 086015 (2018).

\bibitem{Haferkamp}  
J. Haferkamp, P. Faist, N.  B. T. Kothakonda, J. Eisert and N. Y. Halpern,
{\it Linear growth of quantum circuit complexity,}
Nature Physics  {\bf 18}, 528 (2022).
 
 
 \bibitem{Chapman}   
 S. Chapman and  G.  Policastro,
{\it Quantum computational complexity from quantum information to black holes and back,}
 European Physical Journal C {\bf 82}, 128 (2022).
 


 \bibitem{Yo} 
 J. M. Romero, E. Montoya-González and  O. Velazquez-Alvarado,
{\it Quantum entanglement, quantum teleportation, multilinear polynomials and geometry,}
arXiv preprint arXiv:2407.17621, accepted  in International Journal of Geometric Methods in Modern Physics (2025).

 \bibitem{3P1} 
K.  Li, J. Wen, Y. Cai, S. V Ghamsari, C. Li, F. Li,  Z. Zhang, Y. Zhang,  and M. Xiao, 
 {\it  Direct generation of time-energy-entangled W triphotons in atomic vapor,}  Science Advances, 10(37), eado3199 (2024)
 
 \bibitem{3P2} 
Z. Feng, R. Zhuang, S. Liu, G. Liu, K. Li, and  Y. Zhang, 
{\it Observation of optical precursor in time-energy-entangled W triphotons,} Adv. Sci. 12, 2501626 (2025). 

\end{thebibliography}
\end{document}